\documentclass[aps,prl,showpacs,twocolumn,groupedaddress]{revtex4}

\usepackage{graphicx,color}
\usepackage{bm, amsmath, amssymb}

\begin{document}

\title{ Inertial and fluctuational effects on the motion of a Bose superfluid vortex  }

\author{T. Cox$^{1}$ and P.C.E. Stamp$^{1,2}$}

\affiliation{$^{1}$Department of Physics \& Astronomy, University
of British Columbia, Vancouver, BC, Canada V6T 1Z1 \\
$^{2}$Pacific Institute of Theoretical Physics, Vancouver, BC V6T
1Z1, Canada}

\begin{abstract}

We study the motion of a vortex under the influence of a harmonic
force in an approximately two dimensional trapped Bose-condensed
gas. The Hall-Vinen-Iordanskii equations, modified to include a
fluctuational force and an inertial mass term, are solved for the
vortex motion. The mass of the vortex has a strong influence on the
time it takes the vortex to escape the trap. Since the vortex mass
also depends on the trap size we have an additional dependence on
the trap size in the escape time which we compare to the massless
case.
\end{abstract}
\maketitle

\section{Introduction}
Quantum vortices appear in a large range of physical systems ranging
from laboratory condensed matter systems like
superconductors\cite{Blatter94} and neutral superfluids ($^4$He
superfluid\cite{Donnelly91}, superfluid $^3$He, and cold atomic
gases\cite{FetterSvidzinsky}), to exotic excitations playing a role
in the early universe\cite{VilenkinShellard}. Understanding the
motion of these vortices is key to understanding the properties of
such systems.

The motion of superfluid vortices has been studied since the
1950's\cite{HallVinen}. Their dynamics has often been described by
some form of Hall-Vinen-Iordanskii (HVI)
equation\cite{HallVinen,Iordanskii64}, which essentially describes
the different forces acting on a vortex. However until recently the
form and even existence of both the forces\cite{Thouless96,Sonin97}
and the vortex mass \cite{Thouless07} have been controversial.
Recently Thompson and Stamp\cite{ThompsonStamp} derived an equation
of motion for a vortex in a boson superfluid, starting from a low
energy effective field theory for the superfluid. In the limit of
low frequency motion, which turns out to be a semiclassical limit,
this equation of motion reduced to a modified HVI equation, in which
both an inertial term and a Markovian noise term are added to to the
original HVI equation. However it was also found\cite{ThompsonStamp}
that at higher frequencies, strong departures occurred from the HVI
equation, and the vortex coordinate then obeys a different equation
which shows strong retardation effects, with highly non-Markovian
correlations.

In this paper we show how the HVI equations, modified to include the
inertial and noise terms, can be used to discuss the dynamics of a
single vortex in a cold BEC gas. At the temperatures and frequencies
so far employed in experiments on such gases, it turns out we are in
the well within the semiclassical regime. Vortices in trapped
Bose-Einstein condensates were first created in the lab by Matthews
\emph{et al.}\cite{Matthews99}, and since then single vortices and
lattices of vortices in BECs have been studied in various
contexts\cite{Madison00,Anderson00,Raman01,Hadzibabic06}. The motion
of a single vortex in a trapped BEC is the subject of this paper. As
predicted by Rokhsar \cite{Rokhsar97} and observed later observed
experimentally\cite{Matthews99,Anderson00} single vortices created
in an optical trap spiral out from the centre of the trap until they
decay at the edge of the condensate.

There have been fairly extensive theoretical studies of vortex
dynamics in BECs; however these have not included an inertial term.
Jackson \emph{et al.}\cite{Jackson99} and Svidzinsky and Fetter
\cite{SvidzinskyFetter} used the Gross-Pitaevskii equation to
calculate the precession velocity of the vortex. Fedichev and
Shlyaponikov \cite{FedichevShlyaponikov} used a version of the HVI
equations with a term describing dissipation, derived by considering
scattering of the single particle excitations from a stationary
vortex core and a backgroud superflow due to the interaction of the
vortex with the boundary of the condesate. More recently Jackson
\emph{et al.}\cite{Jackson09} performed numerical simulations, and
interpreted the results in terms of an HVI equation with no inertial
mass term or backflow. Duine \emph{et al.} \cite{Duine04} used a
stochastic form of the Gross-Pitaevskii equation at finite
temperature to derive a purely dissipative equation for the motion
of the vortex, including a thermal noise term.

In what follows we study the motion of a vortex in a trap using the
form of the two-dimensional HVI equation derived by Thompson and
Stamp \cite{ThompsonStamp} in the semiclassical limit. We assume the
vortex is in a harmonic potential caused by the trap. We solve these
equations of motion for the motion of the vortex and calculate the
time it takes the vortex to escape the trap. We discuss how the
escape time depends on the radius of the trap and on the
temperature, and we discuss the role that the vortex inertial mass
plays in these results. We also compare the results to previous
work.

%%%%%%%%%%%%%%%%%%%%%%%%%%%%%%%%%%%%%%%%%%%%%%%%%%%%%%%%%%%%%%%%%%%%%%%%%%%%%%%%%%%%%%
%%%%%%%%%%%%%%%%%%%%%%%%%%%%%%%%%%%%%%%%%%%%%%%%%%%%%%%%%%%%%%%%%%%%%%%%%%%%%%%%%%%%%%

\section{Equation of Motion}

A quantum vortex can be described either by looking at the
$N$-particle wave-function of the system, or by defining a vortex
reduced density matrix $\rho({\bf r}, {\bf r'};t) = \langle {\bf r}
| \hat{\rho}(t) | {\bf r'} \rangle$, in which the other degrees of
freedom of the superfluid have been averaged over. If we wish to
find an equation of motion for the vortex itself, we are then
obliged to derive this from the dynamics of $\rho({\bf r}, {\bf
r'};t)$. This density matrix is conveniently rewritten in terms of a
'centre of mass' variable ${\bf R} = ({\bf r} + {\bf r'})/2$ and a
quantum fluctuation variable $\xi = ({\bf r} - {\bf r'})/2$. In the
'classical limit' where the characteristic frequency $\Omega$ of the
vortex dynamics is low (such that $\hbar \Omega \ll kT$, where $T$
is the temperature), the quantum fluctuations $\xi(t)$ become
negligible, and we then expect \cite{ThompsonStamp} a set of
modified HVI equations to be valid for what is now a semiclassical
vortex coordinate ${\bf R}(t)$. In a 2-dimensional Bose superfluid
these take the form
\begin{equation}
 \label{eq:HVI}
M_v\frac{\mathrm{d}^2\mathbf{R}}{\mathrm{d}\, t^2} -\rho\kappa\hat{\mathbf{z}}\times\frac{\mathrm{d}\mathbf{R}}{\mathrm{d}\, t} +
D_0(T)\frac{\mathrm{d}\mathbf{R}}{\mathrm{d}\, t} =\mathbf{F}_{\mathrm{fluc}}(t)+\mathbf{F}_{\mathrm{ap}}(t)
\end{equation}
in a frame of reference where ${\bf v}_n = {\bf v}_s = 0$ (ie.,
botht he normal current and supercurrents are zero; for the more
general case where they are arbitrary, see
ref.\cite{ThompsonStamp}). Here $\mathbf{F}_{\mathrm{ap}}$ is the
applied force on the vortex, $\rho$ is the fluid density,
$\kappa=\frac{h}{m}$ is the quantum of circulation, $D_0(T)$ is the
temperature-dependent longitudinal damping coefficient, and
$\mathbf{F}_{\mathrm{fluc}}(t)$ is a fluctuating force with the high-$T$
Markovian correlator
\begin{equation}
\left\langle F^i_{\mathrm{fluc}}(t)F^j_{\mathrm{fluc}}(s)\right\rangle=\chi(T)\delta^{ij}\delta(t-s)
\end{equation}
in which the angled brackets denote an average over an ensemble of
identically prepared systems. Finally, $M_v$ is the
geometry-dependent vortex mass. For a circular trap geometry with a
radius $R_o \gg a_o$, the vortex core radius, the vortex mass is
\cite{Popov73,Duan94},
\begin{equation}
 \label{eq:mass}
M_v=\pi\rho_sa_o^2\left[\ln\left(\frac{R_o}{a_o}\right)+\gamma_E+1\right]
\end{equation}
where $\rho_s$ is the condensate density and $\gamma_E$ is Euler's constant.

We assume that the vortex is in a harmonic well, so that
\begin{equation}
\mathbf{F}_{\mathrm{ap}}(t)= M_v\omega_o^2 \mathbf{R}(t)  \equiv k_o
\mathbf{R}(t)
\end{equation}
in which $\omega_o$ is the trap frequency, and $k_o$ the 'spring
constant'. In this case the equation of motion can be simplified by
defining the complex position variable ${\mathbb{R}} \equiv
R^x+iR^y$, the complex 'dissipation' $\Gamma\equiv
\frac{D_0}{M_v}-i\frac{\rho\kappa}{M_v}$, a normalized noise
constant $\sigma = \sqrt{\chi}/M_v$, and a normalized complex
fluctuation variable $\xi(t)$, where $\sigma \xi(t) =
[F^x_{\mathrm{fluc}}(t)+iF^y_{\mathrm{fluc}}(t)]$. The equation of
motion then becomes
\begin{equation}
 \label{eq:cplxeom}
\frac{d^2{\mathbb{R}} }{ dt^2}+\Gamma\frac{d {\mathbb{R}} }{ dt} -
M_v\omega_o^2{\mathbb{R}} = \sigma\xi(t).
\end{equation}
This equation is most easily solved by considering the 4-dimensional
phase space position vector $Q$
\begin{equation}
Q(t) = \begin{pmatrix} {\mathbb{\dot{R}}}(t)
\\{\mathbb{R}}(t)\end{pmatrix}.
\end{equation}
In terms of $Q$ the solution to equation \eqref{eq:cplxeom} can be
written,
\begin{equation}
 \label{eq:zeta}
 Q(t)=e^{-\mathcal{M}t}Q(0)+\sigma
e^{-\mathcal{M}t}\int_0^t\mathrm{d}
s\phantom{s}e^{\mathcal{M}s}\begin{pmatrix}\xi(s)\\ 0\end{pmatrix}
\end{equation}
where $\mathcal{M}$ is the matrix
\begin{equation}
\mathcal{M}=\begin{pmatrix}\Gamma&-\omega_o^2\\-1&0\end{pmatrix}.
\end{equation}
From the solution \ref{eq:zeta} we may immediately calculate the
correlator,
\begin{align}
\left\langle Q(t')Q^\dagger(t)\right\rangle =
&e^{-\mathcal{M}t'}\biggl[\left\langle
Q(0)Q^\dagger(0)\right\rangle\biggr.\\
&\biggl.+\; 2\sigma^2\int_0^t\mathrm{d} s
e^{\mathcal{M}s}\begin{pmatrix}1&0\\0&0\end{pmatrix}e^{\mathcal{M}^\dagger
s}\biggr]e^{-\mathcal{M}^\dagger t}
\end{align}
Here a superscript dagger denotes the matrix hermitian conjugate and
it is assumed that $t'\geq t$.

%%%%%%%%%%%%%%%%%%%%%%%%%%%%%%%%%%%%%%%%%%%%%%%%%%%%%%%%%%%%%%%%%%%%%%%%%%%%%%%%%%%%%%
%%%%%%%%%%%%%%%%%%%%%%%%%%%%%%%%%%%%%%%%%%%%%%%%%%%%%%%%%%%%%%%%%%%%%%%%%%%%%%%%%%%%%%

\section{Escape from Trap}

To get an estimate of the time it takes the vortex to escape from
the trap we look at $\left\langle |{\mathbb{R}}(t)|^2\right\rangle$
as a function of time. Since the eigenvalues of $\mathcal{M}$ are
$\frac{1}{2}(\Gamma\pm\Delta)$ where
$\Delta=\sqrt{\Gamma^2+4\omega_o^2}$ is the discriminant of the
characteristic polynomial, we see that the leading order term is for
large $t$ is $\sim \exp\left(2\mathrm{Re}\Delta t \right)$ where
$\mathrm{Re}\Delta$ denotes the real part of $\Delta$. The full leading
order term for $\left\langle |{\mathbb{R}}(t)|^2\right\rangle$ in
the long time limit is
\begin{equation} \label{eq:R2}
\left\langle |{\mathbb{R}}(t)|^2\right\rangle =
\frac{2\sigma^2|\Delta+\Gamma|^2}{|\Delta|^4\left(2\frac{D_0}{M_v} +
\mathrm{Re}(\Delta)\right)}e^{2\mathrm{Re}\Delta t}.
\end{equation}
Note that any contributions from the initial velocity or position of
the vortex are subdominant at large times $t\gg|\Delta|$.

Let us define the 'escape time' $\tau_e$ by setting $\left\langle
|R(\tau_e)|^2\right\rangle=R_o^2$, ie., the time it takes for the
vortex to move to the edge of the circular container; we then find
\begin{eqnarray}
 \label{eq:tau}
\tau_e &=& A\ln\left(\frac{R_o^2}{C}\right) \nonumber \\
&=& \frac{a_o}{4}\sqrt{\frac{\pi\rho_s}{K}}
\left[\ln\left(\frac{R_o}{a_o}\right)\right]^{\frac{1}{2}}
\ln\left(\frac{\sqrt{K^3\pi\rho_s}a_oR_o^2}{\chi}\right).
\end{eqnarray}
where we have defined
\begin{align}
A&=\frac{1}{\mathrm{Re}\Delta}\\
C&=\frac{2\sigma^2|\Delta+\Gamma|^2}{|\Delta|^4\left(2\frac{D_0}{M_v}+\mathrm{Re}\Delta\right)}.
\end{align}
The approximations we have made only give the leading order
dependence of $\tau_e(R_o)$ on $R_o$, for large $R_o$. Three remarks
are in order:

(i) In two dimensions\cite{ThompsonStamp} $\chi\sim T^5$ and
$D_0\sim T^4$, so at low temperature we have
\begin{equation}
\tau_e=A_o\ln\left(\frac{R_o}{BT^5}\right)
\end{equation}
where $B$ is a constant independent of $R_o$ and $A_o = A(T=0)$
which is proportional to $\sqrt{\ln(R_o)}$ for large $R_o$.

(ii) We can also drop the inertial term completely (ie., let $M_v
\rightarrow 0$); this is equivalent to solving the HVI
equations \eqref{eq:HVI} with an added noise term, but no inertial
term. The exact solution for the displacement $\mathbb{R}(t)$ with
the initial conditions $\mathbb{R}(0)=0$ is then
\begin{equation}
\left\langle |\mathbb{R}(t)|^2\right\rangle \;\;\;\underset{\text
M_v \rightarrow 0}{\longrightarrow}\;\;\; \frac{\chi}{k_o D_0}
\left[\exp\left(\frac{2k_oD_0t}{D_0^2+\rho^2\kappa^2}\right)-1\right]
\end{equation}
which leads to the escape time
\begin{equation}
\tau^0_e=\left(\frac{D_0^2+\rho^2\kappa^2}{2k_o D_0}\right)
\ln\left(1+\frac{k_o D_0R_o^2}{\chi}\right)\sim
T^{-4}\ln\left(\frac{R_o^2}{CT}\right)
\end{equation}
where $C$ is a constant. Thus the presence of an inertial term in
the equation of motion leads to a completely different temperature
dependence in the escape time.

 (iii) We can compare these results to the work of
Duine \emph{et al.} \cite{Duine04}, who have $\tau_e\sim
T^{-1}\ln\left(\frac{GR_0^2}{T}\right)$ where $G$ is independent of
both $T$ and $R_o$, and to that of Fedichev and Shlyapnikov
\cite{FedichevShlyaponikov}, who find $\tau_e\sim
T^{-1}\ln\left(\frac{R_o}{R_{\mathbf{min}}}\right)$, where
$R_{\mathbf{min}}$ the initial radial position of the vortex. We see
that our result for $\tau_e$ has a different dependence on both the
trap radius $R_o$ and the temperature $T$. The difference in the
dependence on $R_0$ arises because the inertial mass of the vortex
depends on the trap radius. The difference in the temperature
dependence here, compared to that of Fedichev and Shlyapnikov, comes
(a) because they have no inertial mass term (b) because our
stochastic force (which is absent from their equation) depends
strongly on temperature, and (c) because Fedichev and Shlyapnikov
have a different temperature dependence in $D_0(T)$ from that used
here. While Duine {\it et al.}\cite{Duine04} do have a stochastic
force term in their equation of motion, they have no inertial mass
term.

Finally, we note that the most obvious way in which calculations of
this kind can be tested is in experiments on 'pancake' quasi-2d Bose
condensates, of the kind investigated in MIT and
Paris\cite{Raman01,Hadzibabic06}. In this context, we emphasize that
there are several limitations to our calculation. First, we have
ignored the effect of the boundary of the condensate. Fetter and
Svidzinsky\cite{FetterSvidzinsky} introduced a background superflow
to the equation of motion to cancel the normal component of the
superfluid at the boundary of the condensate, and this will
presumably affect our result. Second, the Thompson-Stamp
derivation\cite{ThompsonStamp} assumed that local deviations from
the mean superfluid density were small. This assumption does not
always hold for Bose-Einstein condensates, and it will be
interesting to see how the results may be modified for BECs in the
extreme compressible limit. Finally, we note that at sufficiently
low temperatures, we expect serious departures from the modified HVI
equations, coming from the non-local terms found by Thompson and
Stamp - experimental probes of this regime will be of great
interest, particularly in view of the long-standing controversy over
the correct equations of motion for a quantum
vortex\cite{Thouless96,Sonin97,Thouless07}.

\section{Conclusions}
Using the modified form for the HVI equations that was found by
Thompson and Stamp in the low-frequency semiclassical
limit\cite{ThompsonStamp}, we have studied the motion of a
superfluid vortex in a harmonic trap. We calculated the life-time of
the vortex in a trap for this case, and find that the temperature
dependence of the trap escape time is different than that found in
previous calculations\cite{Duine04,FetterSvidzinsky}, even when the
inertia of the vortex is neglected. The vortex inertia also has a
significant influence on this escape time. Experimental tests of
results like this will be possible on trapped cold BECs, and will
allow significant tests of the theory of vortex dynamics (which have
been very difficult until now).

\begin{acknowledgements}

This work was supported by NSERC, by CIFAR,  and by PITP.

\end{acknowledgements}

%\pagebreak

\end{document}